\documentclass{aa}
\usepackage{graphicx}

\begin{document}
\title{The electron temperature of the inner halo of the Planetary Nebula 
\object{NGC~6543}}

\author{Siek Hyung\inst{1}\and Garrelt Mellema\inst{2} \and 
  Seong-Jae Lee\inst{3} \and Hyouk Kim\inst{4}}

\institute{Korea Astronomy Observatory, 61-1 Whaam-dong, Yusong-gu, Taejon 305-348,  Korea\\
              \email{hyung@kao.re.kr} \and
 Sterrewacht Leiden, P.O. Box 9513, 2300~RA, Leiden, the Netherlands\\
              \email{mellema@strw.leidenuniv.nl} 
\and 
 Dept. of Astronomy and Space Science, Chungnam National University, 220 Kung-dong, Yusong-gu, Taejon 305-764, Korea\\
              \email{seong@canopus.chungnam.ac.kr}
\and 
Dept. of Earth Science Education, Korea National University of Education, Chung-Buk 363-791, Korea \\
              \email{vitkim@kao.re.kr}
}  

\offprints{Garrelt Mellema, \email{mellema@strw.leidenuniv.nl}}

\date{Received 14 June 2001 / Accepted 21 August 2001}

\abstract{We investigate the electron temperature of the inner halo and
nebular core regions of \object{NGC~6543}, using archival Hubble Space Telescope ({\it
HST}) Wide Field Planetary Camera 2 ({\it WFPC2}) images taken through
narrow band [\ion{O}{iii}] filters. Balick et al.~(2001) showed that the
inner halo consists of a number of spherical shells. We find the temperature
of this inner halo to be much higher ($\sim$15\,000~K) than that of the
bright core nebula ($\sim$8500~K).  Photo-ionization models indicate that
hardening of the UV radiation from the central star cannot be the main source
of the higher temperature in the halo region. Using a radiation hydrodynamic
simulation, we show that mass loss and velocity variations in the AGB wind
can explain the observed shells, as well as the higher electron temperature.
 }

\maketitle

\section{Introduction}
\object{NGC~6543} (\object{PN G096.4+29.9}) is one of the more intriguing Planetary Nebulae
(PNe) because of its complex morphology. The core nebula (the ``Cat's Eye'')
consists of several shells and rings projected on top of each other. Images
taken with the Chandra X-ray telescope have for the first time revealed the
central wind-driven bubble (Chu et al.~2001). Along the symmetry axis, the
PN shows FLIERs (Fast Low Ionization Emission Regions).

The object is surrounded by a `giant halo' of 50\arcmin, which has an
overall spherical morphology, but { which} at the outer edge consists of many
irregularly shaped gas clouds (Middlemass, Clegg, Walsh 1989, henceforth
MCW89).  This giant halo has a low expansion velocity of at most
5~km~s$^{-1}$, and despite {its} chaotic appearance, {it} does not show any distinct
kinematic features (Bryce et al.~1992). One of the brighter filaments is
characterized by a high electron temperature of 15\,000~K, {derived from 
an analysis of} the
[\ion{O}{iii}] line ratio (MCW89), but by a much lower temperature of 9000~K when
considering the H$\alpha$ and [\ion{N}{ii}]6584\AA\ profiles (Meaburn et al.~1991).

Recently, Balick et al.~(2001) (henceforth BWH01) analyzed archival Hubble
Space Telescope ({\it HST}) images and found that the region just outside the
bright core shows at least nine circular `rings', {which are} in fact
projected spherical shells. We will refer to this region as the ``inner
halo'' region to distinguish it from the ``outer halo'' studied by MCW89 and
Meaburn et al.~({1991}). This inner halo looks much more continuous and
regular than the outer halo. Rings like the ones seen around \object{NGC~6543} are
quite common, and are found around AGB stars like \object{IRC~10216} (Mauron
\& Huggins 2000), post-AGB stars like \object{CRL~2688} (Sahai et al.~1998),
and other PNe, such as \object{NGC~7027}, and \object{Hb~5}. The time scale
associated with the implied mass loss variations are typically 100--1000
years, which is too long to associate them with the pulsations of AGB stars,
and too short to connect them to thermal pulses. They could originate from
variations inherent to the nature of the mass loss mechanism during the AGB
{phase} (Simis et al.~2001), or from stellar magnetic field variations
({Soker 2000;} Garcia-Segura et al.~2001), {or from the influence of
a companion star (Harpaz et al.~1997; Mastrodemos \& Morris 1999).}

The central star of \object{NGC~6543} is a Wolf-Rayet star of spectral type
[WC~8]. The P Cygni profiles of, for example, \ion{C}{iv} and \ion{N}{iv} in
its spectrum indicate a terminal wind velocity of $V(\infty) =
2150$~km~s$^{-1}$ (Castor et al.~1981).  The chemical abundances in
\object{NGC~6543} are likely to be position dependent (Hyung et
al.~2000). Its distance is not well known; values derived through statistical
methods are around 1~kpc, and according to other methods it could be as close
as 500~pc. We will assume a distance of 1~kpc.

In this paper, we investigate the inner halo region by deriving the
[\ion{O}{iii}]5007/4363 line ratio from archival narrow band {\it HST}\/
images, using a proper interstellar extinction correction. This procedure and
the {results} are described in Sect.~2. In Sect.~3 we attempt to {explain} the
results using both photo-ionization and radiation-hydrodynamic
modelling. We discuss the {results} in Sect.~4, and {summarize} our conclusions in
Sect.~5.

\section{\textit{HST/WFPC2}\/ Data}
The {\it HST}\/ archive images are available from the {\it HST}\/ archival
data centre.  The [\ion{O}{iii}]4363 and 5007 emission-line images can be
used to find an [\ion{O}{iii}] electron temperature map.  We searched the
archive for {\it WFPC2}\/ images of [\ion{O}{iii}] and H$\alpha$/H$\beta$.
As will be explained below, we need an H$\alpha$ or H$\beta$ image to
eliminate the continuum contribution from the [\ion{O}{iii}] images. The {\it
HST}\/ archive contains 11 images of \object{NGC~6543} in these filters. After careful
examination, we selected the three best {ones}: U27Q0103T ([\ion{O}{iii}]4363);
U27Q010FT (H$\alpha$); and U27Q010AT ([\ion{O}{iii}]5007).  The central part
of the [\ion{O}{iii}]4363 image is saturated by the central star.  See
Table~1 for the details.

\begin{table}
{Table 1 -- {\it WFPC2}\/ Archive Images}\\[1ex]
\label{tbl-1}
\begin{tabular*}{8cm}{llll}
\hline\hline
Name & Filter & [Ion] line & Exp Time (s)\\
\hline
  U27Q0103T & F437N & [\ion{O}{iii}]4363 & 1200\\   
  U27Q010AT & F502N & [\ion{O}{iii}]5007 & 600\\
  U27Q010FT & F656N & H$\alpha$   & 200\\
\hline
\end{tabular*}\\[0.5ex]
{{All these observations were performed on September 18, 1994.}}
\end{table}

\subsection{Reduction of the {\it WFPC2}\/ Images}

The archived {\it WFPC2}\/ images have already been calibrated via the pipeline
procedure, i.e.\ the analog-to-digital correction, bias level removal, bias
image subtraction, dark image subtraction, flat-field multiplication, and
shutter shading correction.  Using standard IRAF routines, we removed {the} cosmic
rays. The retrieved images were converted into flux units (or magnitudes)
using the header information.

\subsubsection{Filter bandwidth correction}

The width and transmission efficiency of the three filters differ (see
Table 2). To compare the images, one must correct for these differences.
We use a normalizing function to {obtain  the observed photon flux within
a bandwidth of 20\AA}: 
\begin{equation}
   F_\mathrm{norm} = F_{HST} / Q(\mathrm{20\AA}), 
\end{equation}
where $F_{HST}$ is the calibrated final flux; $Q$(20\AA) (given in the
last column of Table~2) is the transmission efficiency for an ideal 20\AA\
bandwidth filter with 100\% transmission.  Once $F_\mathrm{norm}$ has been
obtained for each image, we {may} compare them with each other.

\begin{table}
{Table 2 -- {\it WFPC2}\/ Narrow Band Filter Set}\\[1ex]
\label{tbl-2}
\begin{tabular*}{8cm}{llll}
\hline\hline
Name (Ion) & Width (\AA) & Peak($\lambda$)\footnotemark[1] & $Q$(20\AA)\footnotemark[2]\\
\hline
  F437N([\ion{O}{iii}]) & 25.2      & 49.6(4368)          & 86.4\%  \\
  F502N([\ion{O}{iii}]) & 26.9      & 57.9(5009)          & 105.0\%  \\
  F656N(H$\alpha$)    & 21.9      & 86.2(6561)          & 112.3\%  \\
\hline
\end{tabular*}\\[0.5ex]
$^1${\footnotesize Peak transmission efficiency in \% ({wavelength $\lambda$ at peak)}}.\\
$^2${\footnotesize The transmitted light relative to an ideal 20\AA\ bandwidth filter,
 assuming this ideal filter has {100\% transmission over a with 20\AA\ bandwidth.}}
\end{table}

\subsubsection{Interstellar extinction} 
The interstellar extinction is
corrected using the relation

\begin{equation} 
F_\mathrm{e-c}  = F_\mathrm{norm} \times 10^{C \cdot k_{\lambda}},
\end{equation}

where $C$ is the logarithmic extinction coefficient at H$\beta$, and
$k_{\lambda}$ the extinction parameter from Seaton (1979). 
{$C$(H$\beta$) values of 0.18, 0.20 and 0.30 have been presented by Kaler
(1976), Middlemass et al. (1989), and Hyung et al. (2000), respectively.
In what follows we will adopt $C$(H$\beta$)=0.20.  The uncertainty in
the adopted $C$ value does not affect the results derived in this paper.} 

\begin{figure*}
{\includegraphics[width=8cm]{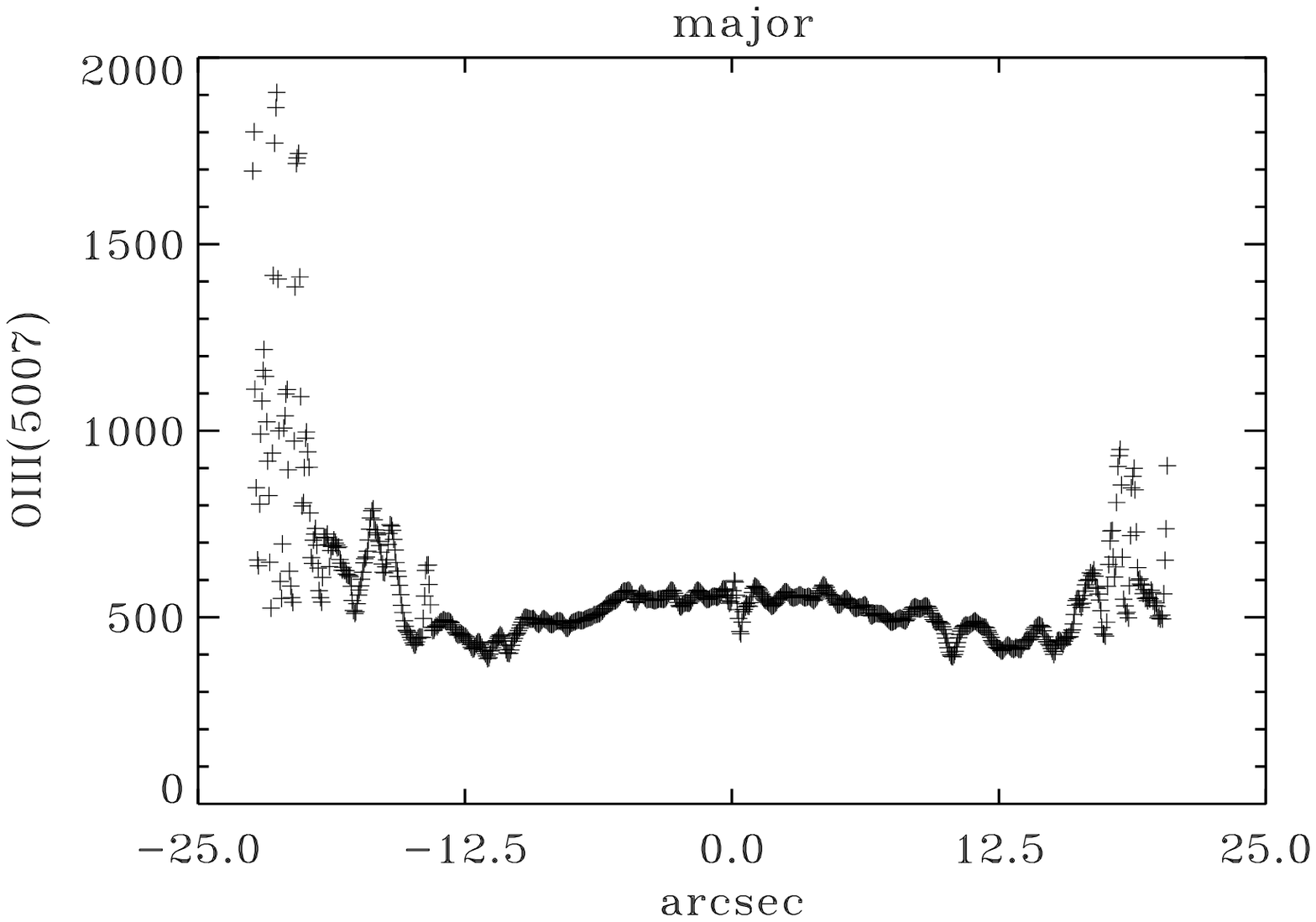}}
{\includegraphics[width=8cm]{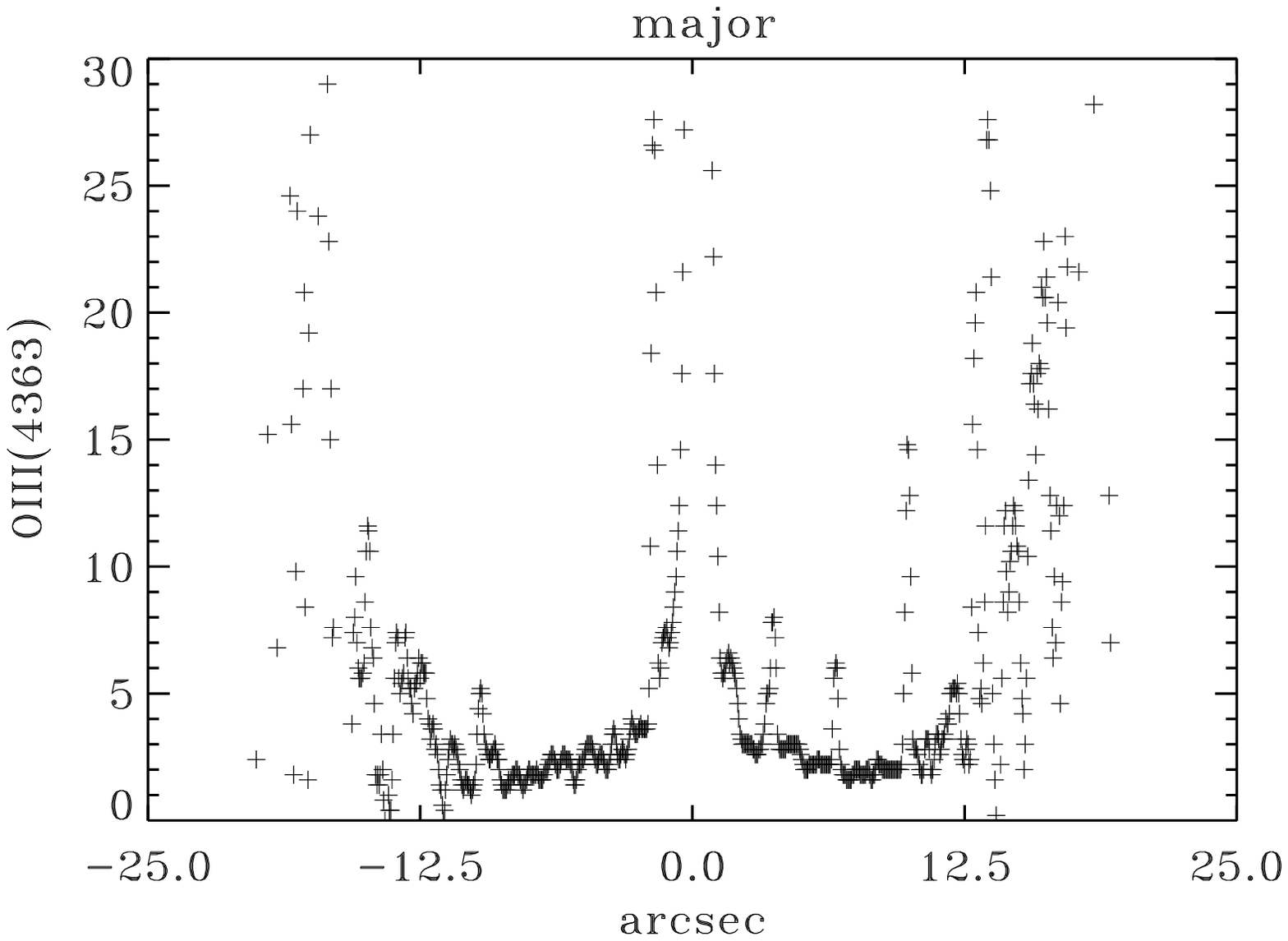}}
\caption{[\ion{O}{iii}] intensities along the major axis [scaled to I(H$\beta$)
= 100]. On the left I([\ion{O}{iii}]5007), on the right I([\ion{O}{iii}]4363).}
\label{oiii}
\end{figure*}

\subsubsection{Continuum correction} 

The filter widths are $\sim$20\AA, while the full width at half maximum
(FWHM) of the emission lines is only $\sim$1\AA.  Thus, the images can contain a
significant contribution from continuum photons. Since no offset nebular filter
images are available in the {\it HST}\/ archive, and the continuum is mostly
produced by hydrogen quanta (see below), we used the H$\alpha$ {\it WFPC2}\/
image to correct for the continuum contribution in the [\ion{O}{iii}] filter images.

First, we estimated the contribution to the continuum of H recombination plus
bremsstrahlung, assuming an electron temperature of 10\,000~K and He/H
abundance ratios of $N$(He$^+$)/$N$(H$^+) = 0.1$ and $N$(He$^{++}$)/$N$(H$^+)
=0.01$.  We find this contribution to be 92.7\% near  [\ion{O}{iii}]4363
emission line, and 91.2\% near  [\ion{O}{iii}]5007.  The theoretical
continuum-emission coefficients have been adopted from Brown \& Mathews
(1970); see their Tables~1--5.  This simple calculation shows that hydrogen
emission dominates the continuum at the wavelengths of interest.

The continuum-corrected [\ion{O}{iii}] images can {then} be obtained from the following
relations,

\begin{equation}
F_\mathrm{f}(5007) = F_\mathrm{e-c}(5007) - 
F_\mathrm{e-c}(\mathrm{H\alpha}) \times A(5007), 
\end{equation}

\begin{equation}
F_\mathrm{f}(4363) = F_\mathrm{e-c}(4363) - 
F_\mathrm{e-c}(\mathrm{H\alpha})  \times B(4363), 
\end{equation}

where $A$(5007) and $B$(4363) are the extinction-corrected continuum fluxes
near 5007\AA\ and 4363\AA, respectively, divided by the extinction-corrected
continuum flux near the H$\alpha$ line + the H$\alpha$ emission flux
itself. Values for $A$ and $B$ were derived from our ground-based Hamilton
Echelle Spectrograph ({\it HES}) data (Hyung et al. 2000): $A$(5007) $\simeq
1.21\times 10^{-3}$ and $B$(4363) $\simeq 5.28 \times 10^{-3}$. {Aller
(1984) lists the theoretical values as $A$(5007) = $3.1 \sim$ $4.4 \times
10^{-3}$ and $B$(4363) = $3.5 \sim 6.2$ $\times 10^{-3}$, and also shows that
$B$(4363)/$A$(5007) $\sim 1.2$.  Our $B$ value agrees with this within 20\%,
but we find a 3 to 4 times smaller value for $A$(5007). This deviation is
probably caused by an overcorrection for scattered light in analysis of the
{\it HES} data.  As a test we also derived the values for the bright core
nebula, and there our procedure does reproduce the expected theoretical
values. However, the strength of the [\ion{O}{iii}]5007 is such that an error
of a factor of four in $A$(5007), only affects the derived intensity to less
than 0.5\%. An error of about 20\% in $B$(4363) also does not influence our
results.}

Although the ratio of continuum to line emission may have positional
variations, we assume that $A$ and $B$ remain constant throughout the nebula
and halo.  This assumption is unavoidable, due to the unavailability of
offset continuum filter images.  It may be not a bad assumption, since the
stratification effect of the H$\alpha$ (or H$\beta$) recombination image is
generally not conspicuous, compared with that of the line images,
e.g.~[\ion{O}{iii}]5007 or [\ion{N}{ii}]6548. The F437N image may suffer from
some light contamination by H$\gamma$ at 4341\AA.

In order to reduce the scatter, we rebinned the image from 0\farcs
1$\times$0\farcs 1 to 0\farcs 3$\times$0\farcs 3 pixels. Figure~\ref{oiii}
displays the [\ion{O}{iii}] 5007 and 4363 line intensities on a flux scale
where I(H$\beta$)=100.  Note how the [\ion{O}{iii}]4363 intensities within
$\sim 2\arcsec$ of the centre are strongly affected by the saturation {caused
by} the central star.

\begin{table}
{Table 3 -- [\ion{O}{iii}] line intensities}\\ 
\begin{tabular*}{8cm}{*{4}{c}} 
\hline\hline
Name(Ions)    &   Core (North)$^1$    &  Core (East)\footnotemark[1]  & Halo\footnotemark[2]  \\
\hline
[\ion{O}{iii}]4363 &  2.05   & 1.58 & 28   \\
{[}\ion{O}{iii}{]}5007 &  704   & 518 & 1528  \\
\hline
\end{tabular*}\\
$^1${\footnotesize Hyung et al. (2000)}\\
$^2${\footnotesize MCW89}\\
[0.5ex]
{Note: Extinction corrected intensities are given based on I(H$\beta$~4861) = 100
($C = 0.2$, 0.3 for {\it HST}\/ and {\it HES}, respectively). See the main text.} 
\end{table}

Table~3 lists the spectroscopic line intensities of the core H~II region
(Hyung et al.~2000) and the outer halo (MCW89). A comparison
with Fig.~\ref{oiii} shows that, within the uncertainties, the
correspondence between the spectroscopic data and the images is reasonable.

\subsubsection{[\ion{O}{iii}] temperature map}

\begin{figure}
{\includegraphics[width=7cm]{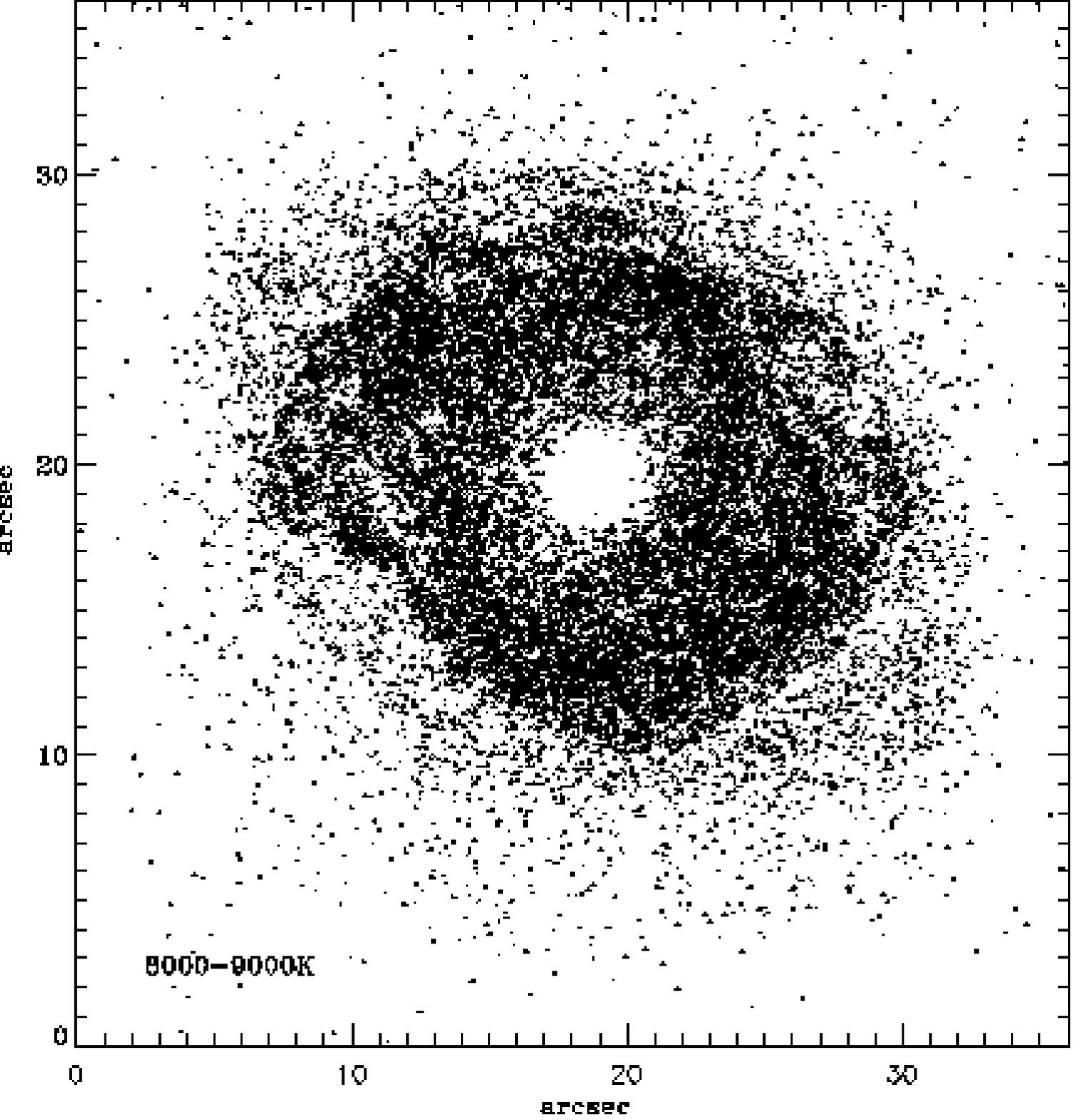}}
{\includegraphics[width=7cm]{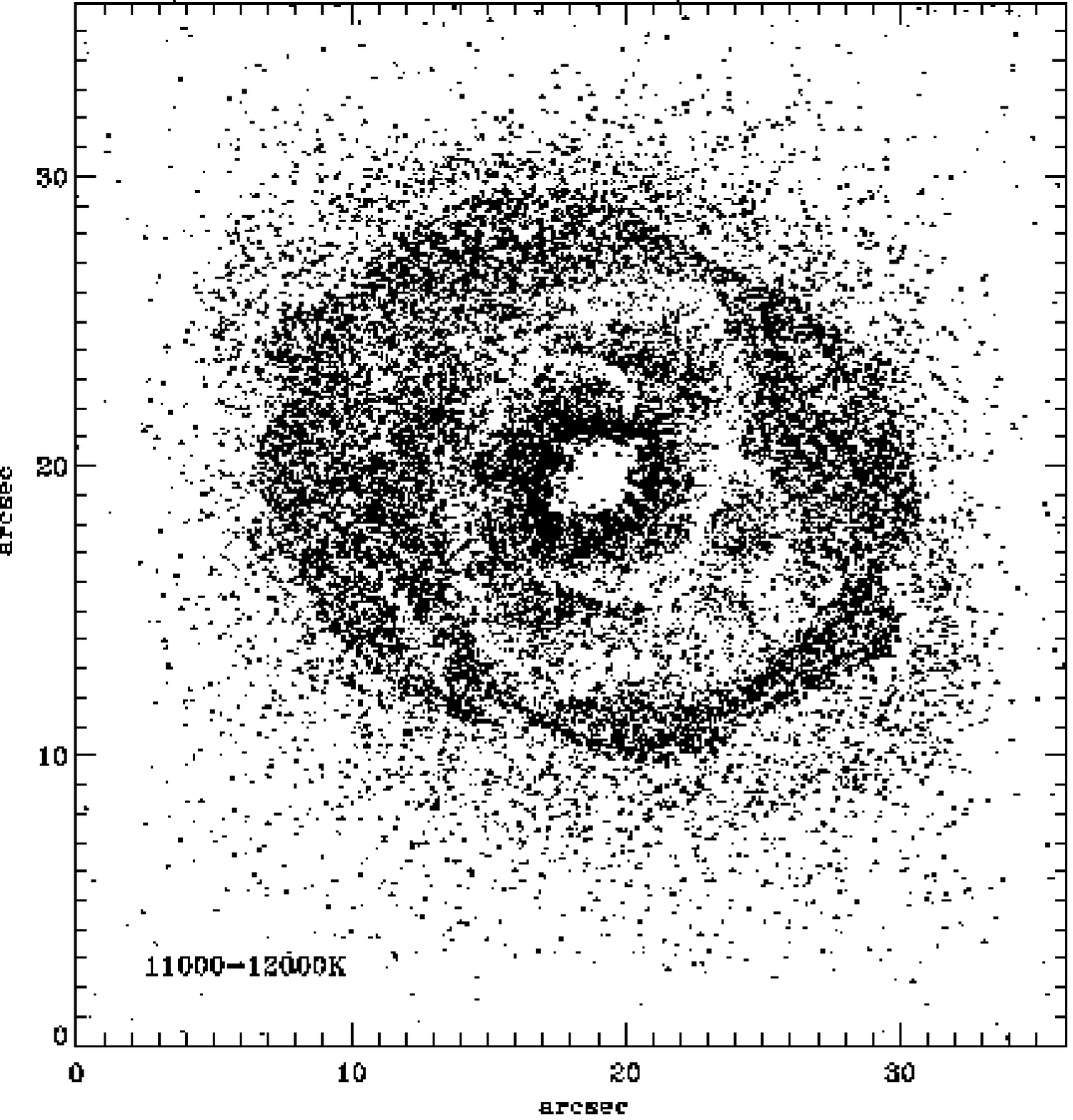}}
{\includegraphics[width=7cm]{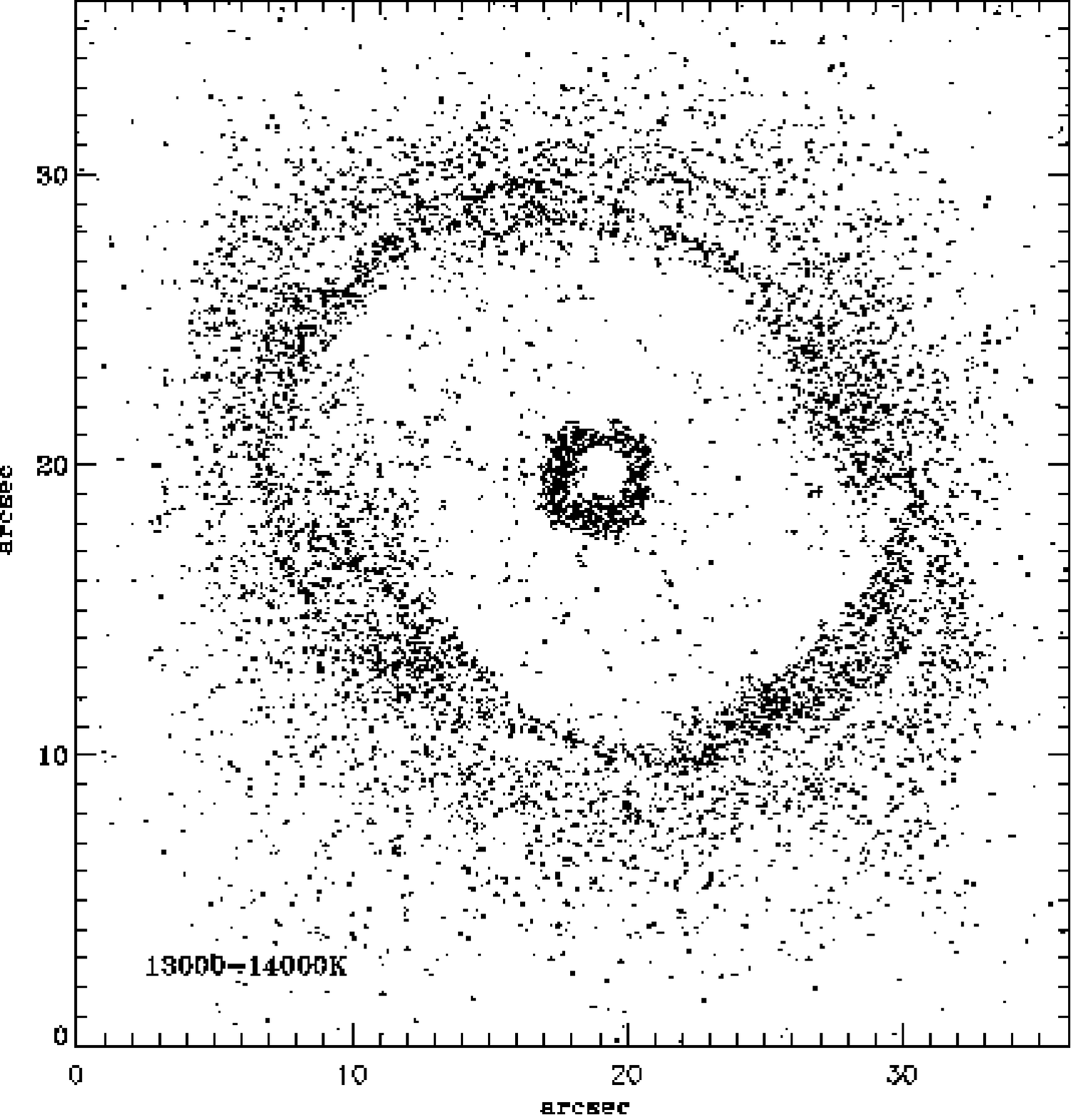}}
\caption{[\ion{O}{iii}] temperature maps for the ranges {\bf a)}
T$_\mathrm{e}([\ion{O}{iii}])$ = 8000 $\sim$ 9000~K, {\bf b)} T$_\mathrm{e}([\ion{O}{iii}])$ {11\,000 $\sim$ 12\,000~K}, {\bf c)} T$_\mathrm{e}([\ion{O}{iii}])$ {13\,000 $\sim$ 14\,000~K}. An upper limit of
20\,000~K has been set for the area around the CSPN.}
\label{tempmaps}
\end{figure}

Figure~\ref{tempmaps} shows the temperature maps for three different electron
temperatures ranges: T$_\mathrm{e} = 8000 - 9000$~K; 11\,000 $-$ 12\,000~K; and
13\,000 $-$ 14\,000~K, respectively. As was found in previous ground-based
studies, the electron temperature of the {core nebula} is 8000--9000~K.  The
$11\,000 - 12\,000~$K image clearly displays an ellipsoidal envelope
{surrounding the core nebula}. Figure~\ref{tempmaps}c shows that the inner halo
region contains hotter material of T$_\mathrm{e}\sim$14\,000~K.  The temperature
in the very central region also appears to be high. However, those values are
{suspect}, due to the saturation by the central star in the [\ion{O}{iii}]4363
image.

\begin{figure}
{\includegraphics[width=8cm]{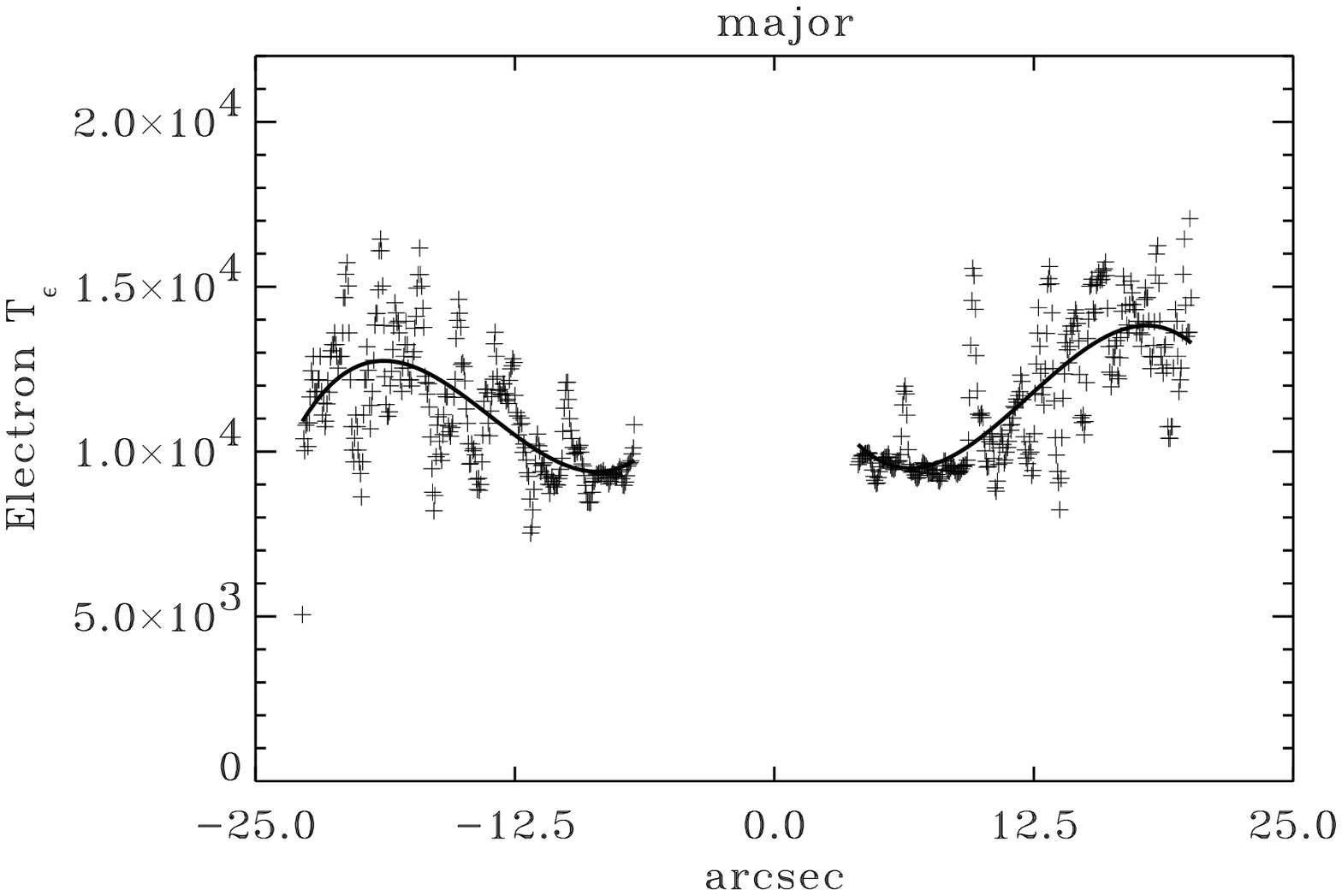}}
\vskip 2mm
{\includegraphics[width=8cm]{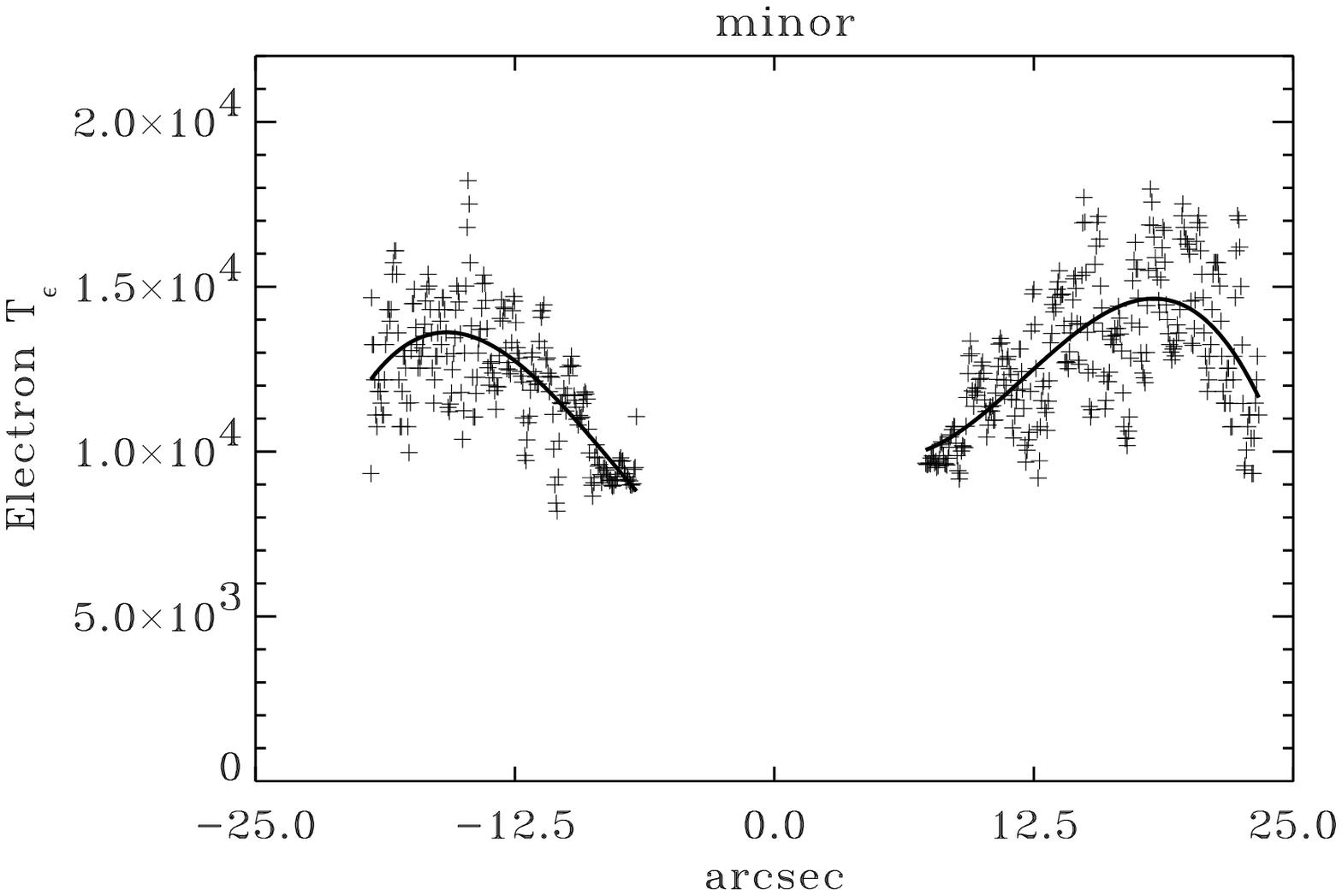}}
\vskip 2mm
{\includegraphics[width=8cm]{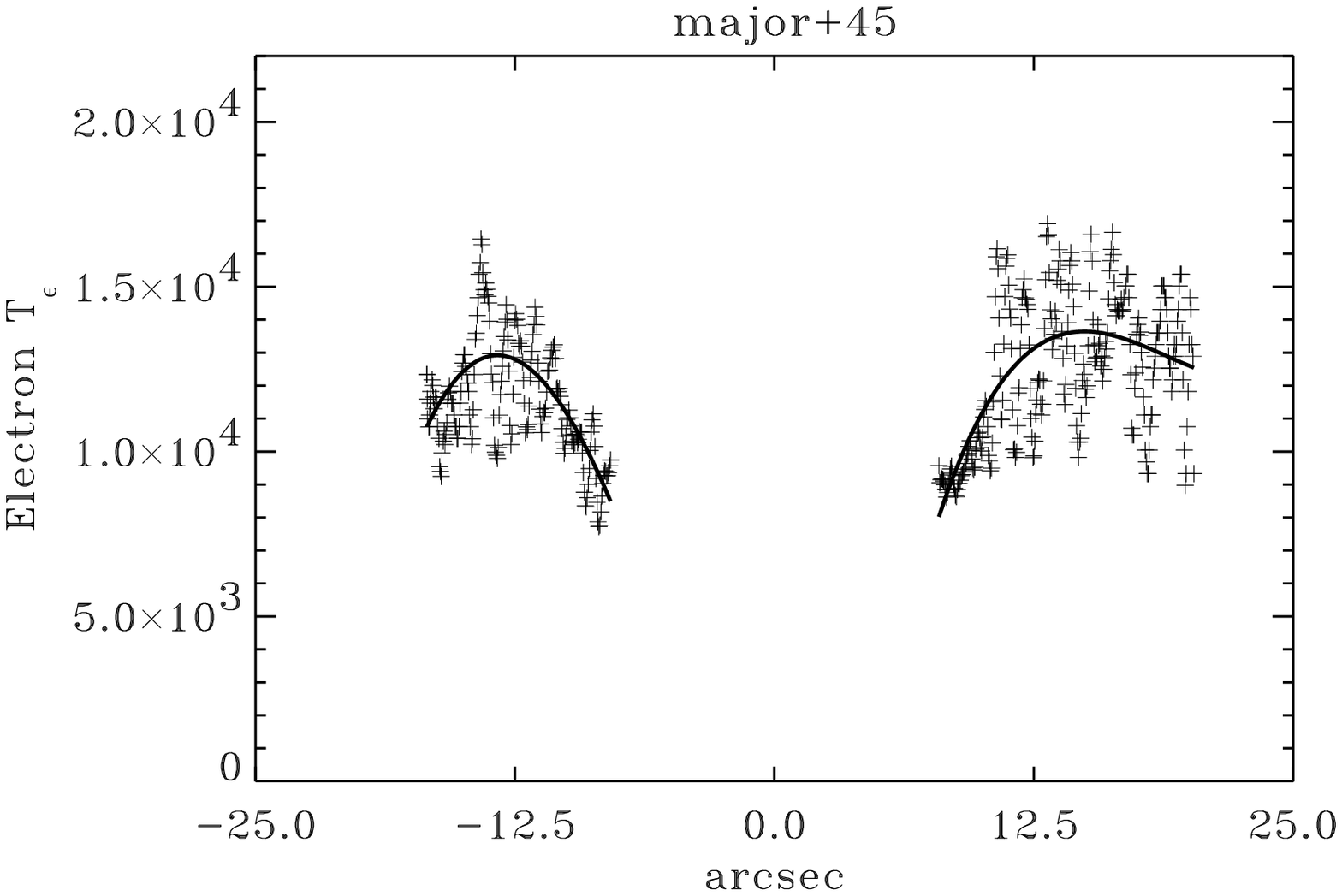}}
\caption{[\ion{O}{iii}] electron temperatures in K: From top to bottom: along
the major axis, along the minor axis, and in between the major and minor axes
(at major + 45$^{\circ}$).  Since binning of 3$\times 3$ pixels was applied,
the spatial resolution is 0\farcs 3. We exclude the region near the CSPN.}
\label{tempgraphs}
\end{figure}

In Fig.~\ref{tempgraphs}, we plot the electron temperature distribution along
the major and minor axes, and at an {intermediate}  angle. The position angles are
$-32^{\circ}$, +58$^{\circ}$, and +13$^{\circ}$, respectively.

We estimated the errors in Figs.~\ref{tempmaps} and \ref{tempgraphs} using a
qualitative analysis. We found them to be less than $\triangle$T$_\mathrm{e}
\sim500$~K in the core nebula, and at least three times larger in the inner
halo, $\simeq 1500$~K.  This latter region could be partly filled with
relatively low temperature gas, like that seen in Fig.~\ref{tempmaps}a. In
this case, the scatter in Fig.~\ref{tempgraphs} may reflect real temperature
fluctuations involving geometrical irregularities. Otherwise, it would be due
to errors of around 3000~K in the halo.

Figure~\ref{tempgraphs} shows a number of {local peaks} with
FWHM${\sim}0\farcs 3$--$0\farcs 4$, where the electron temperature increases
sharply.  These could be due to blobs or filaments. We compared temperature
profiles, e.g.\ $0\farcs 8$ apart, parallel to the major and minor axes. No
structure survived in the next adjacent scan; apparently, blobs substantially
larger than $0\farcs 4$ do not exist. This leads to the conclusion that any
existing substructures must be about $0\farcs 4$ (0.002~pc for an assumed
distance of 1 kpc), or less.

Figure \ref{tempgraphs} also shows third order Legendre polynomial {curves
fitted} to the radial electron temperature distribution (excluding the central
region).  These fits suggest that the maximum electron temperature in the
halo occurs at a radial distance of $r\sim$ 0.09pc (18\arcsec, 17\arcsec, and
14\arcsec, respectively), beyond which it {falls off} again.  The radial
distribution of the {\it HST}\/ electron temperature suggested by the
polynomial fitting indicates a high electron temperature location, which may
roughly form a complete circle of $r\sim$ 0.09~pc.  The halo temperature
fitting found from the {\it HST}\/ images may be the projection of
a relatively large shell, of $r\sim$0.09--0.1~pc.

\section{Excitation by Hardened UV or Shock?}

The thermal properties of the gas in the halo are set by photo-ionization
heating, and also possibly by shock heating. We will now investigate whether
{these two processes can explain the observed electron temperatures}.

\subsection{Hardening of the UV radiation field}
Long ago, Aller, Baker and Menzel (1939) suggested that a hardening of the UV
radiation field from the central star might be responsible for a rise of
$T_\mathrm{e}$ in the outer part of photoionized PNe.  Photoionization modeling
by Hyung et al. (2000) suggested that the electron temperature near the outer
boundary (at {0.056~pc}) of the core H~II region of \object{NGC~6543} is higher than that
of the inner region (at {0.05~pc}) by 600~K, due to a hardening of far UV
stellar radiation.  Can this hardened stellar UV radiation also raise the
electron temperature of the halo gas to the value suggested by the {\it
HST}\/ [\ion{O}{iii}] images?  MCW89 showed that the emission in
the outer halo could not be produced by photo-electric heating: they were
only able to reproduce a halo temperature some 2000~K above the core region
temperature.

We extend the photoionization model by Hyung et al. (2000) to see whether the
hardened UV radiation from the WR-type central star produces the high
electron temperatures in the inner halo, or gives the observed [\ion{O}{iii}]
5007/4363 line ratios.  Note that the photoionization model by Hyung et
al.~(2000) indicates that \object{NGC~6543}'s bright core is density bounded.  Thus,
there are still plenty of hard, far-UV photons available to ionize the gas in
the halo region.  Since the relatively low density (35 cm$^{-3}$) outer halo
case was already tested by MCW89, we explore higher
densities for the inner halo gas. The description of the photoionization
model, the improved Sobolev approximation (ISA) model atmosphere for the
hydrogen deficient Wolf Rayet type central star of the planetary nebula (CSPN),
and references to selected atomic parameters may be found in Hyung et
al.~(2000).  We adopted the abundances from the core nebula, {as
described by}  Hyung
et al.~(2000). The results show that a higher electron temperature can only
be achieved for very high densities. For, example, the model with $N_\mathrm{H}$ =
10$^5$~cm$^{-3}$ and a small filling factor, $f = 0.0001$, gives
I(4959+5007)/I(4363) = 105.8 (corresponding to $T_\mathrm{e} \sim$
12\,000~K).  Since the spectroscopic observations indicate that the electron
density in the halo is low, this high density model is not an appropriate
solution to the halo emission.  Nonetheless, there may  be some implications
for the physical conditions of the filaments observed in the central or halo
region of some PNe -- the filaments are density enhanced structures, in
any case.

\subsection{Shock heating}
An alternative way to heat gas is through shocks. Shocks are caused either by
pressure or velocity differences. In order to produce a noticeable
temperature increase of the order of 15\,000~K, the shock speed needs to be
at least 20~km~s$^{-1}$. One origin for such shocks would be a `leaky'
stellar wind bubble. If the material which the fast wind is running into is
quite clumpy, a mass-loaded flow would proceed beyond the core nebula into
the halo region, where it would interact with halo material to produce the
necessary shocks. This is the model favoured by Meaburn et al.~(1991) to
explain the measured electron temperatures in the outer halo. Whereas the
structures in the outer halo also morphologically suggest such an
interaction, the inner halo looks much smoother, and shows no signs of
interacting with a mass-loaded flow from the core. In fact, with the
parameters given in Meaburn et al.~(1991), the density in the inner halo
would be dominated by the mass-loaded flow. {In support of the mass-loaded
flow model, we should mention} that the images presented in BWH01 do reveal
radial structures in the inner halo, which mostly (but not uniquely) appear
to originate from the FLIER regions in the core nebula. If these are
extensions of a faster outflow in the FLIERs, and not illumination effects,
they could certainly produce (local) shocks in the inner halo.

On the other hand, the presence of the rings in the inner halo region already
indicates that there are {density and hence} pressure differences there. The idea that the inner
halo is dynamic is strengthened by the increase in the line widths in the
inner halo region to about 30~km~s$^{-1}$ (Bryce et al.~1992; BWH01), which
could be explained by a superposition of faster moving shells. In this case
the shells would have velocities well in excess of the 5~km~s$^{-1}$ derived
for the outer halo. If the shells were created while the CSPN was on the AGB,
velocity differences would also have helped in their survival up to the
current era. Density variations without velocity variations tend to be
smoothed out.  If the shells are produced by the mechanism explained in Simis
et al.~(2001), velocity differences of 50~km~s$^{-1}$ or higher can be
reached (Simis, private communication).

It will be difficult to measure the true expansion velocity of the
shells, since the slit position which would capture it the best, would also
contain the bright and kinematically complicated core nebula (see Miranda \&
Solf 1992).

\begin{figure}
{\includegraphics[width=8cm]{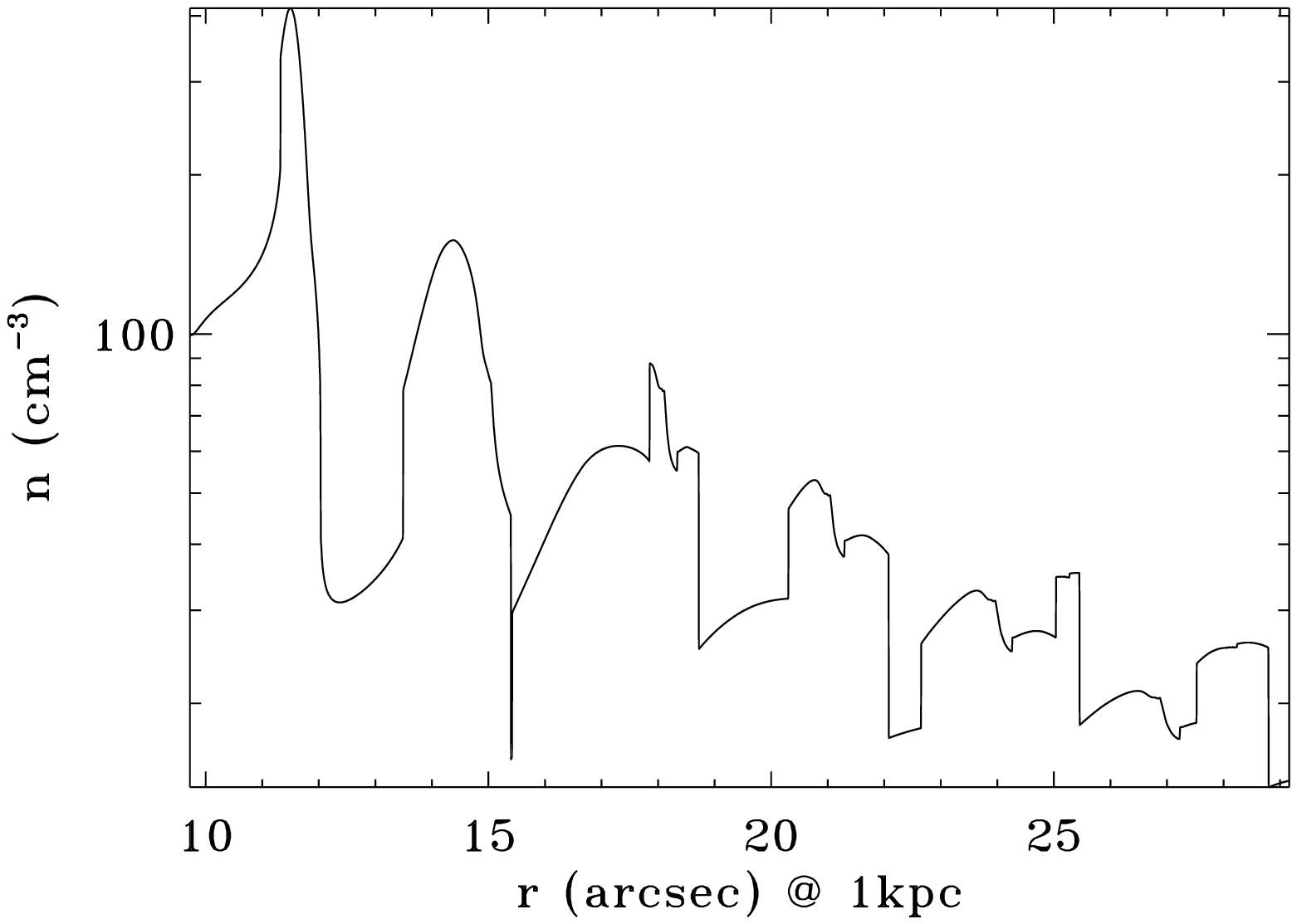}}
{\includegraphics[width=8cm]{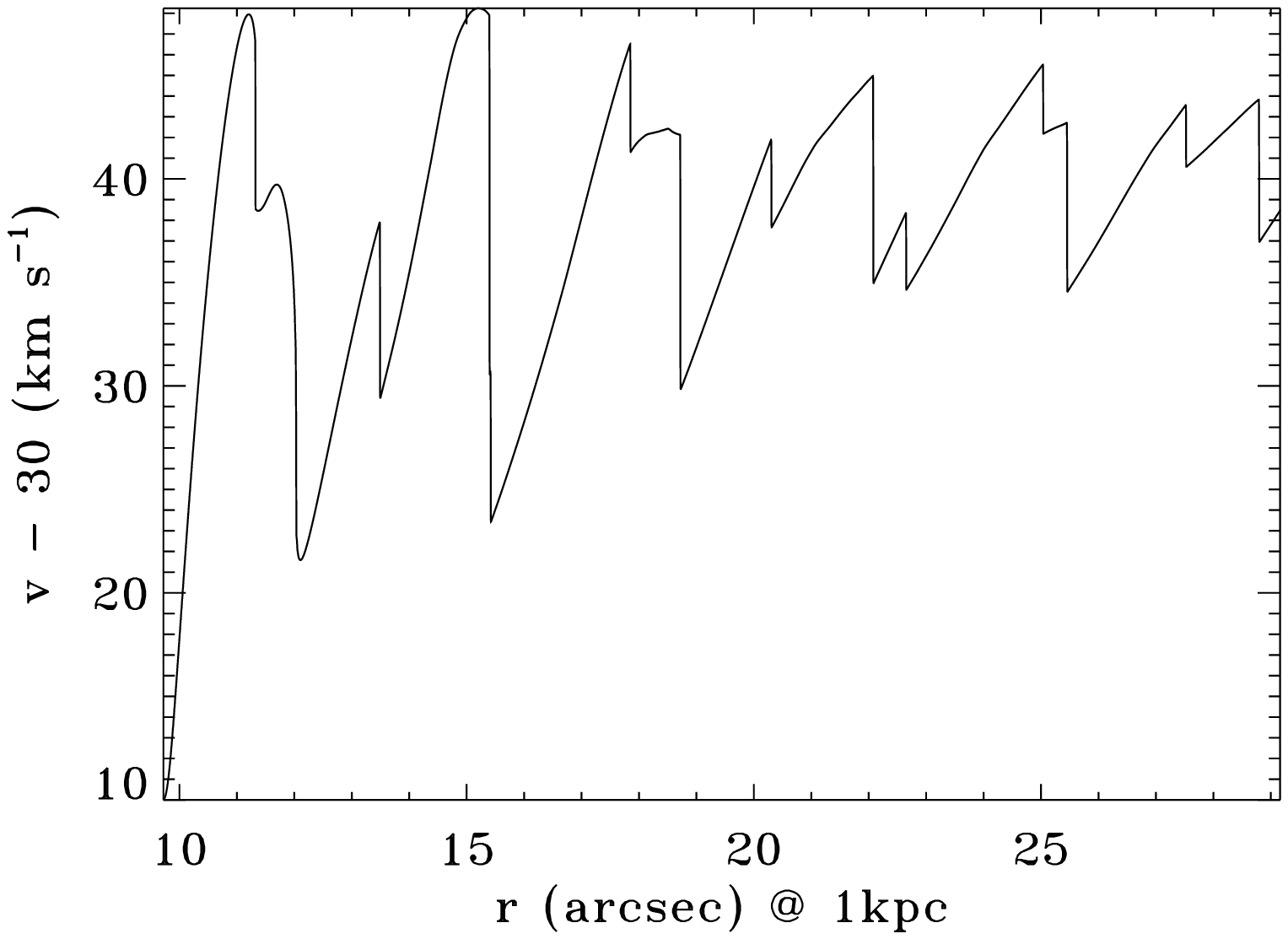}}
{\includegraphics[width=8cm]{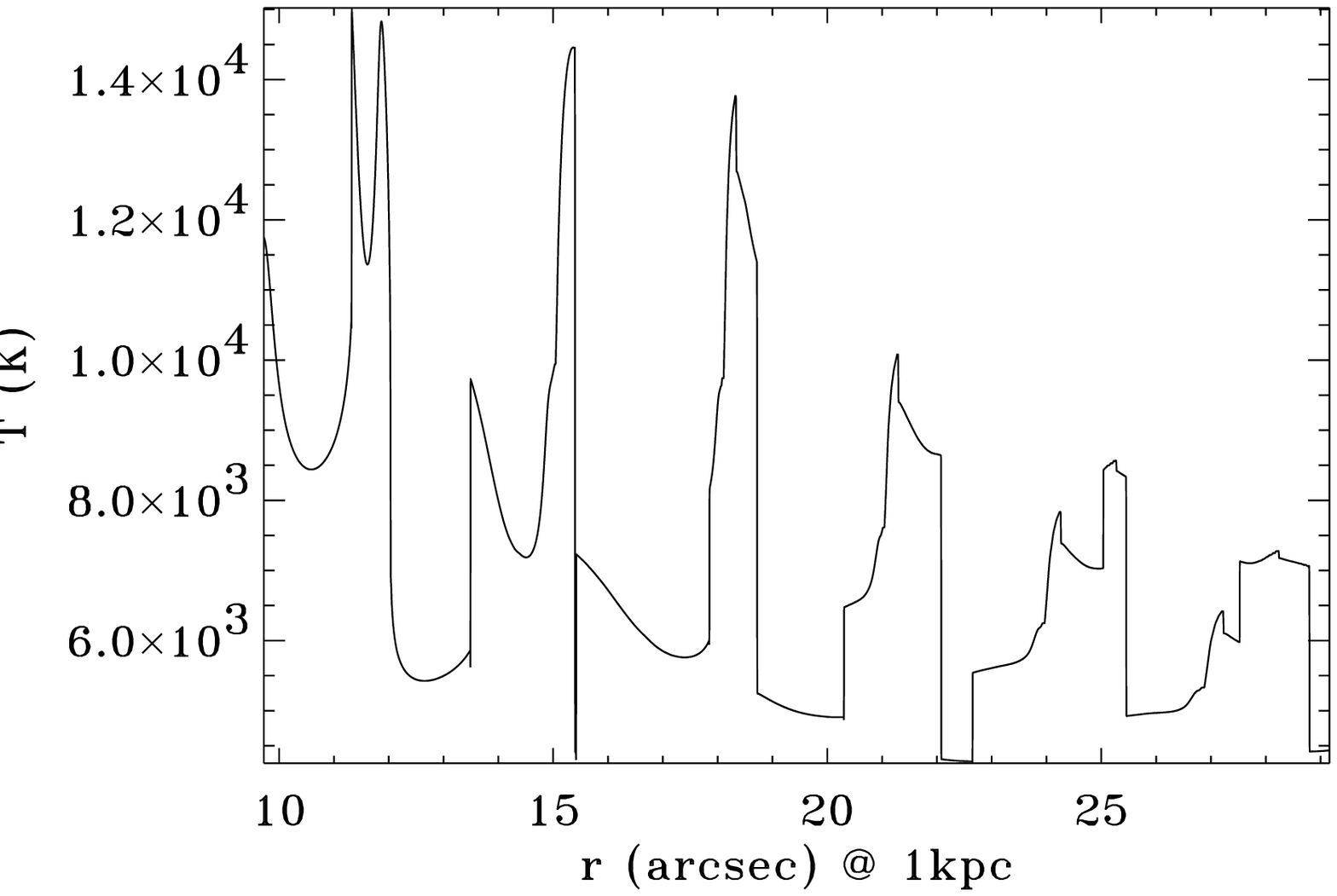}}
\caption{Hydrodynamics simulation. Graphs showing the number density ($n$), velocity ($v$), and temperature ($T$). The time is 2550~years.}\label{dvt-sim}
\end{figure}
 
\subsubsection{Numerical model}
In order to show the feasibility of shock heating by the shells, we set up a
simple numerical experiment. For this we use a hydrodynamic code coupled to a
photo-ionization calculation. The hydrodynamic part of the code is based on
Roe's approximate Riemann solver (Mellema 1993; Eulderink \& Mellema 1995). In each
numerical time step,  the heating and cooling rates are calculated using the
{\it DORIC}\/ routines (Raga et al.~1997; Mellema et al.~1998), which
calculate the non-equilibrium ionization, heating and cooling of the
gas. Since we are not concerned with the shaping of the entire nebula, and
want to resolve the post-shock cooling region, we consider only the
one-dimensional (spherical) case. For this we used 12\,000 computational cells
of size $2 \times 10^{11}$~cm. At this cell size, the
Courant-Friedrichs-Lewis (CFL) time step imposed by the hydrodynamics is
much smaller than the cooling time, and all cooling regions are well resolved.

\begin{table}
{Table 4 -- Input parameters for the simulation}\\
\begin{tabular*}{8cm}{*{2}{l}}
\hline\hline
\.{M}$_\mathrm{low}$          &  $4\times 10^{-6}$~M$_{\odot}$ yr$^{-1}$  \\
$v_\mathrm{low}$           &  40 km~s$^{-1}$   \\
\.{M}$_\mathrm{high}$         &  $2\times 10^{-5}$~M$_{\odot}$ yr$^{-1}$  \\
$v_\mathrm{high}$          &  80 km~s$^{-1}$   \\
$\tau_\mathrm{variation}$     &  200 years \\
$r_\mathrm{in}$ (boundary)    &  0.05 pc (10\arcsec) \\
$r_\mathrm{out}$ (boundary)   &  0.15 pc (30\arcsec) \\
$\Delta r$                 &  6.48 $\mu$pc \\
Central Star $T_\mathrm{eff}$ &  48\,000 K \\   
Central Star $L$           &  3510 L$_{\odot}$   \\ 
\hline
\end{tabular*}
\end{table}

The optical radius of the main core is about 10\arcsec, while the halo
region shown in the {\it HST}\/ images corresponds to an area with a radius
of about 30\arcsec, which is 0.15 pc.  Thus, we set the inner radius of the
computational grid to be 0.05~pc, and let the grid run out to 0.15 pc.

At the start of a simulation, the value of the electron temperature is given
by the equilibrium temperature provided by the photo-ionization model.

For the abundances, we used the values derived by Hyung et al.~(2000) for the
core nebula: He/H = 0.13, C/H = 5.0(-5), N/H = 1.2(-4), O/H = 3.0(-4), Ne/H =
5.0(-5), S/H = 7.0(-6).

Bryce et al.~(1992) found the outer halo to be kinematically inert, with an
upper limit to the expansion velocity of 4.6 km~s$^{-1}$.  The observed
electron density of the halo region obtained from the [O~II] diagnostics can
set a constraint for the mass loss rate of the slow wind.  Middlemass found
35$^{100}_{-35}$ cm$^{-3}$ (in the outer halo, outside the {\it HST}\/
image). Taking a (pre-shock) number density of $n_\mathrm{H} \sim$20
cm$^{-3}$ at 0.15~pc, we arrive at a mass loss rate for the slow wind of
\.{M}$_\mathrm{slow}$ = 10$^{-6}$~M$_{\odot}$ yr$^{-1}$.  Alternatively,
BWH01 find a total mass of $\sim 0.1$~M$_\odot$ ejected between 17,500 and
6500 years ago (assuming a slow wind velocity of 10~km~s$^{-1}$), which gives
a mass loss rate of $\sim 10^{-5}$~M$_{\odot}$ yr$^{-1}$. For a slower wind
velocity of 5~km~s$^{-1}$, one obtains half this value. We take a base slow
wind mass loss rate of $10^{-6}$~M$_{\odot}$~yr$^{-1}$.

At the inner edge of the grid, we then vary the mass loss rate between this
base value and 10$^{-5}$~M$_{\odot}$ yr$^{-1}$, and with a velocity difference of
40~km~s$^{-1}$. The variation is sinusoidal with a period of 200~years.

\begin{figure*}
\centerline{\includegraphics[width=8cm]{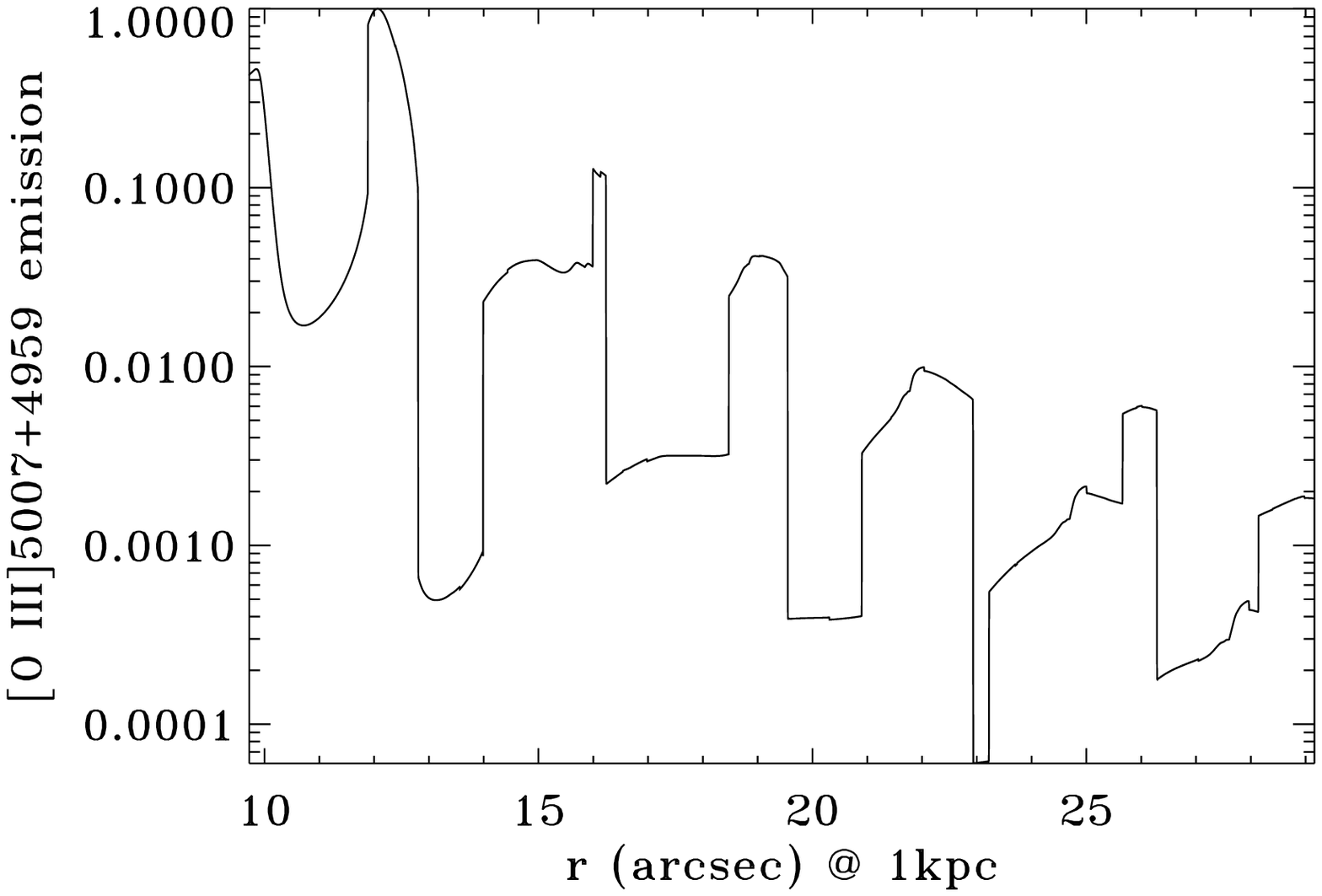}
\includegraphics[width=8cm]{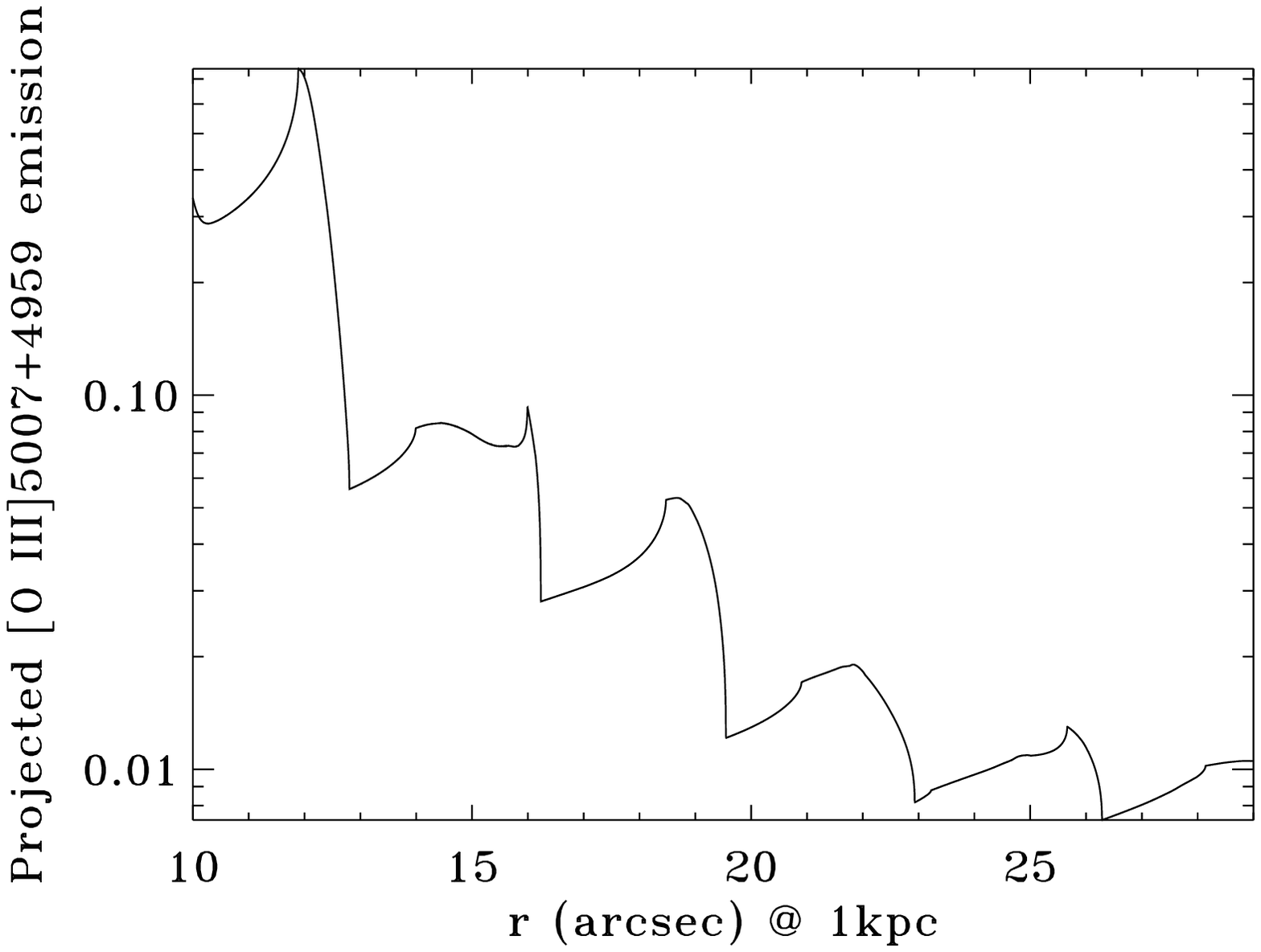}}
\centerline{\includegraphics[width=8cm]{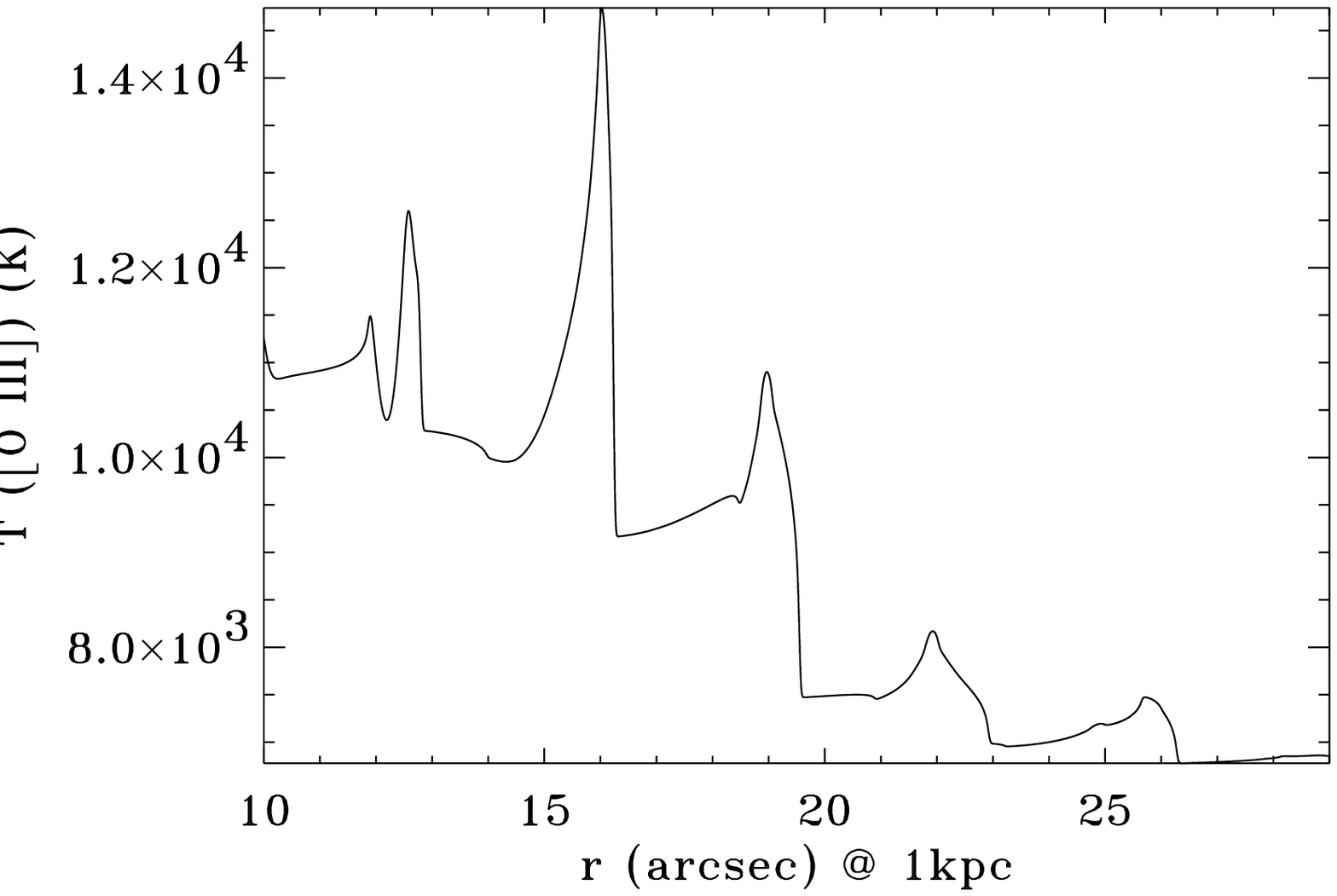}
\includegraphics[width=8cm]{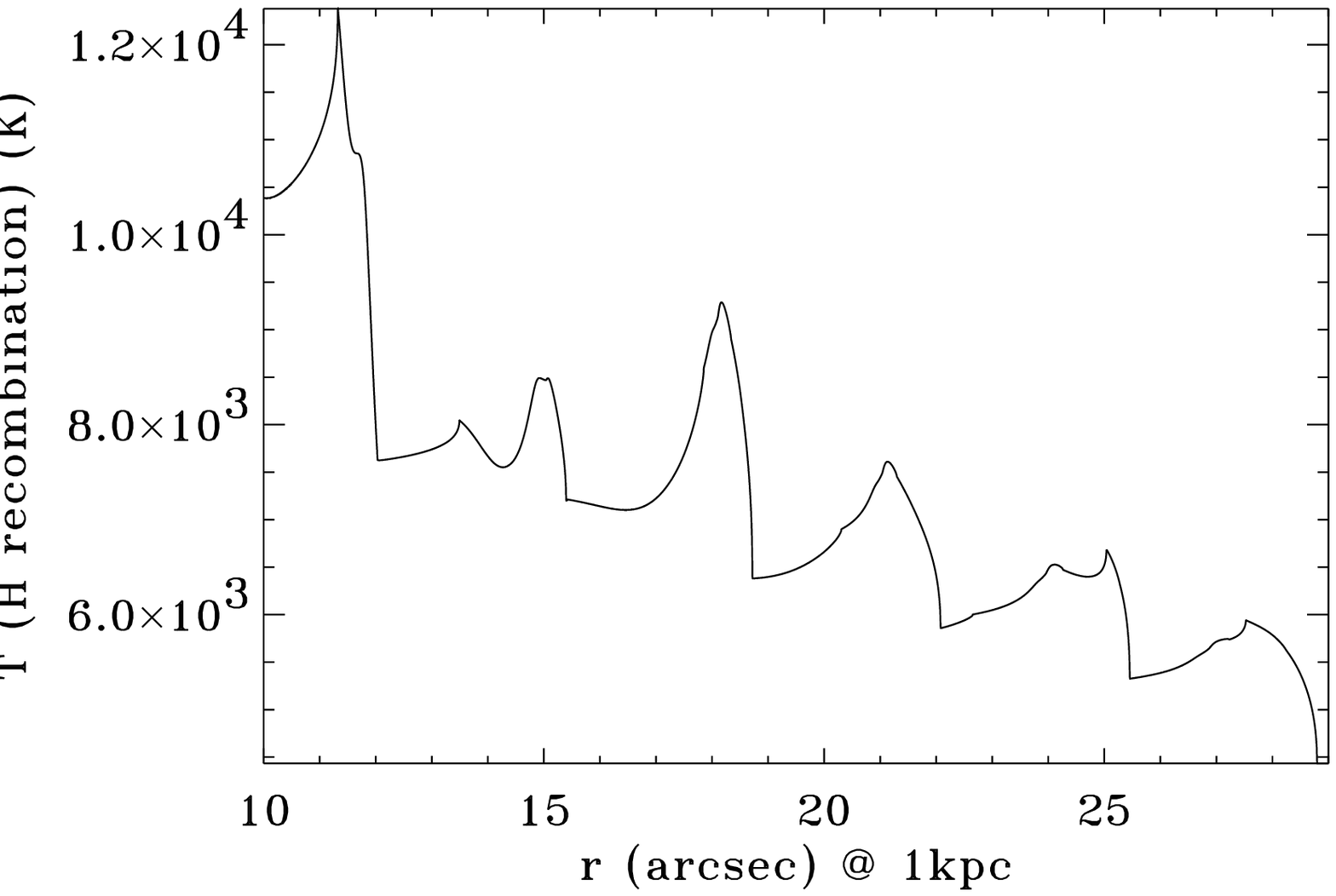}}
\caption{Top left: [\ion{O}{iii}](5007+4959) intensity as a function of
radius, in arbitrary units. Top right: the same but projected on the sky.
Bottom left: the electron temperature derived from the
[\ion{O}{iii}](5007+4959)/4363. Bottom right: the electron temperature
derived from H recombination; also compare these temperatures to
Fig.~\ref{dvt-sim}c.}\label{oiii-sim}
\end{figure*}
 
The typical sound speed in an ionized gas is around 10~km~s$^{-1}$. This
means that if we choose a base inflow velocity of 5 or 10~km~s$^{-1}$, the
inflow will be {close to}  sonic. In this case a steady wind will quickly
accelerate (similar to a Parker wind solution) and achieve supersonic
velocities. In practice this means that a fixed inner boundary inflow
velocity of around 10~km~s$^{-1}$ cannot be used. In previous simulations
(see e.g.~Mellema 1994), this was never a problem, since the inflow velocity
was the highly supersonic fast wind. Here, where we want to concentrate on
the halo, this physical behaviour poses a problem. To get around this, we
decided to use inflow velocities which are well over sonic. In practice this
means a base inflow velocity of 40~km~s$^{-1}$, and consequently a peak
inflow velocity of 80~km~s$^{-1}$. We scaled the mass loss rates up by a
factor of 4 and 2, to achieve roughly the same densities as before.

For the photo-ionization calculation we assumed the column densities at the
inner grid edge to be $10^{12}$ cm$^{-2}$ for \ion{H}{i} and \ion{He}{i}, and
$10^{15}$~cm$^{-2}$ for \ion{He}{ii}. The parameters adopted in the
simulation are listed in Table~4.

\begin{figure*}
\centerline{\includegraphics[width=8cm]{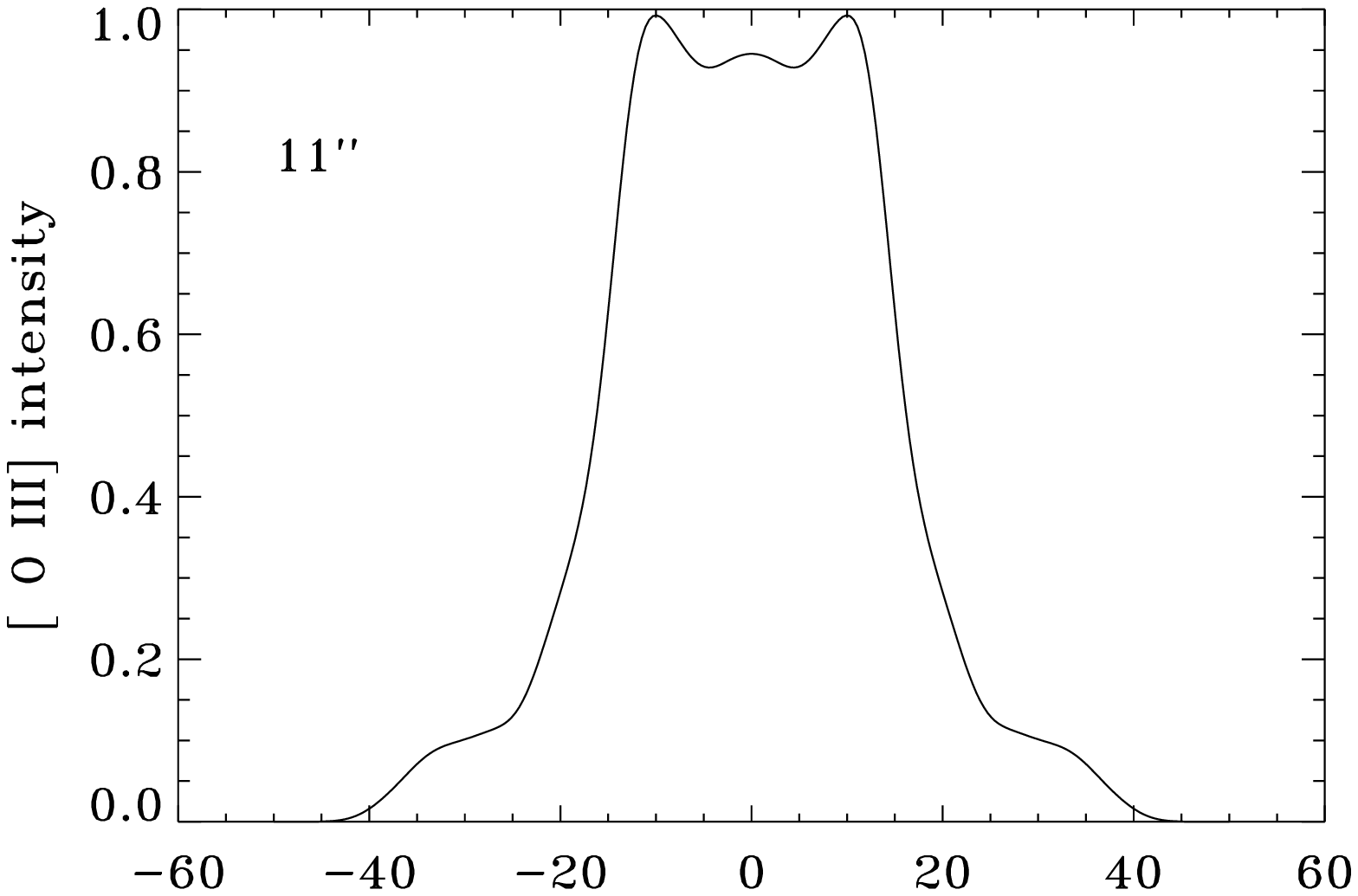}
\includegraphics[width=8cm]{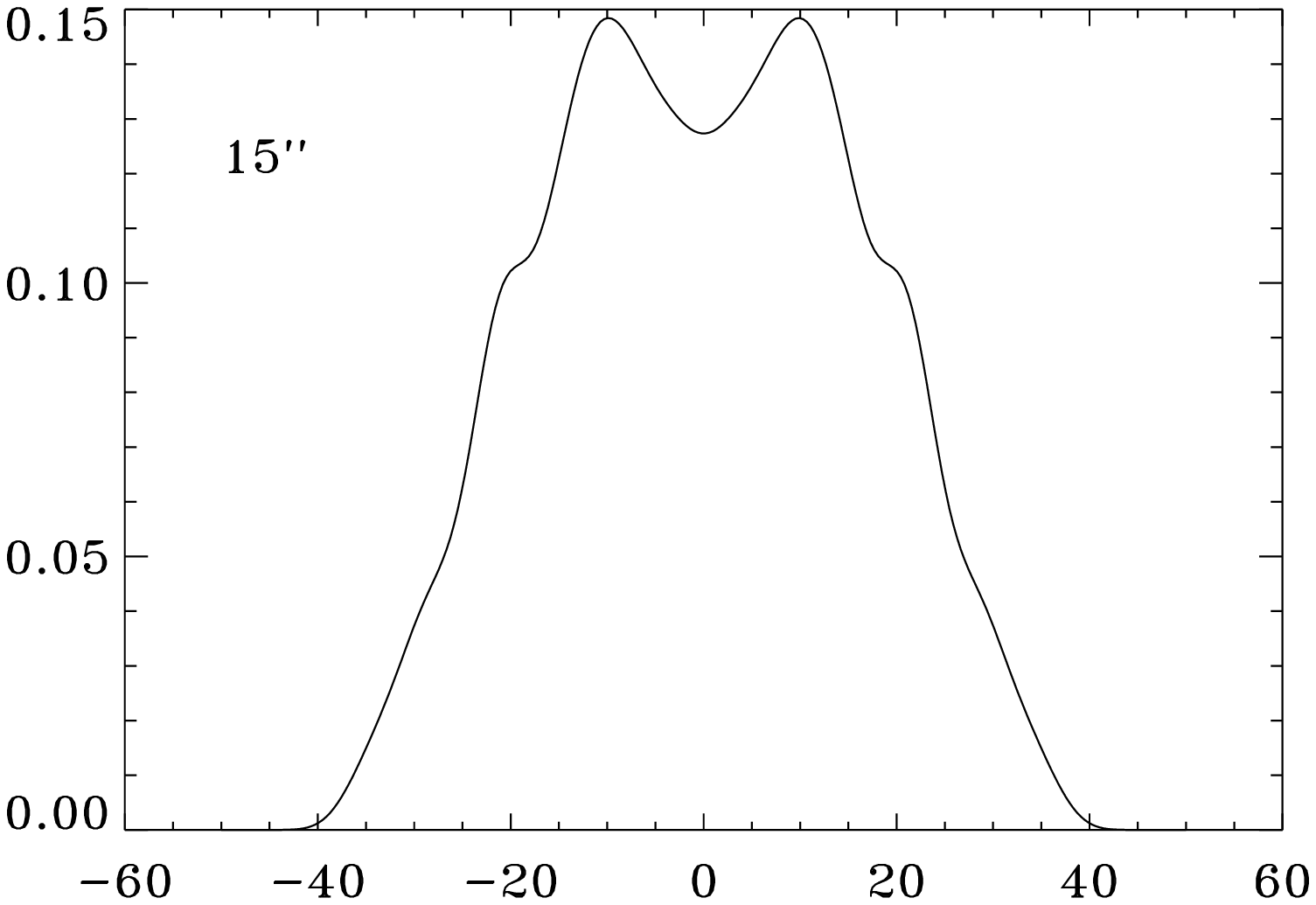}}
\centerline{\includegraphics[width=8cm]{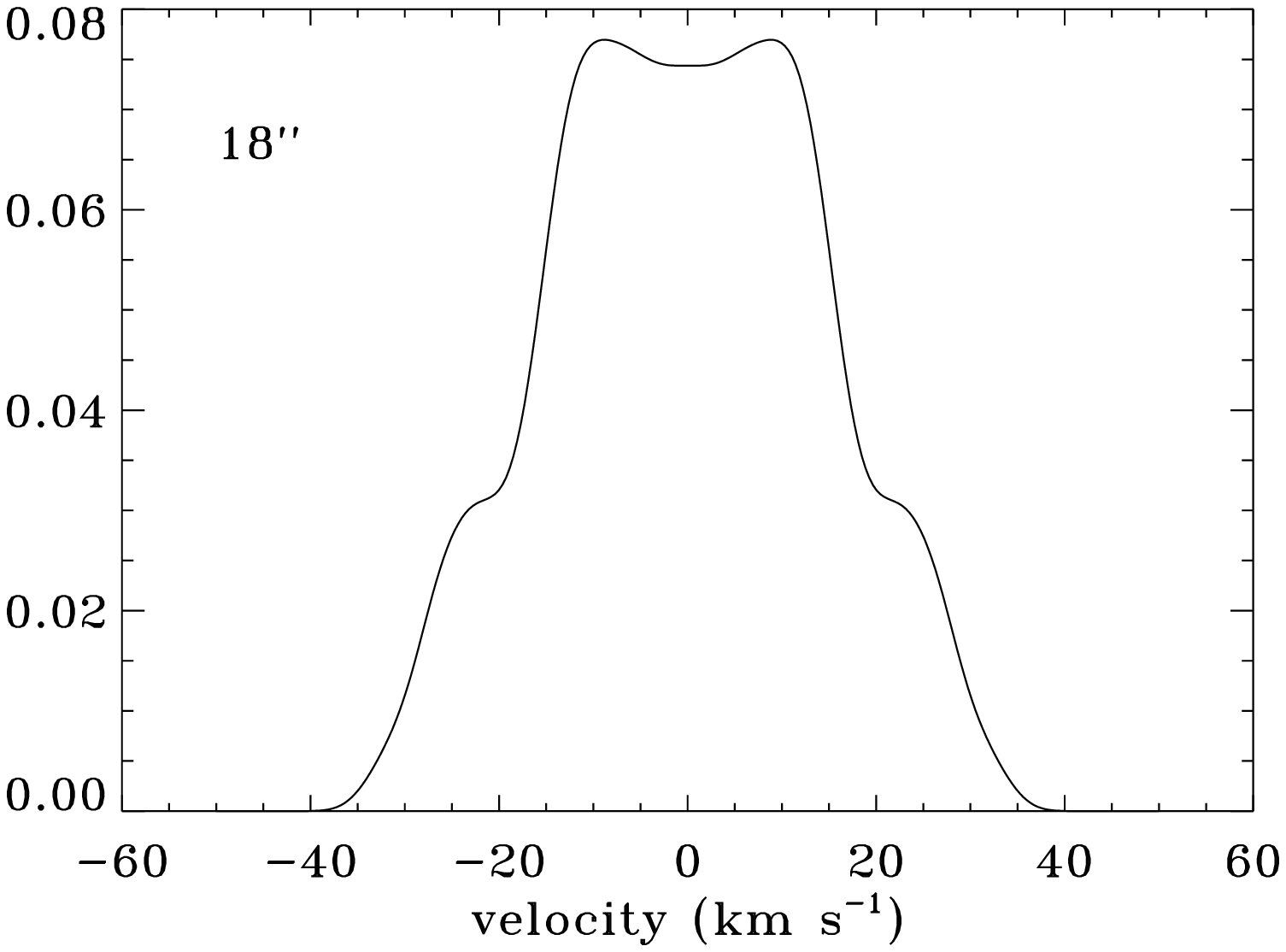}
\includegraphics[width=8cm]{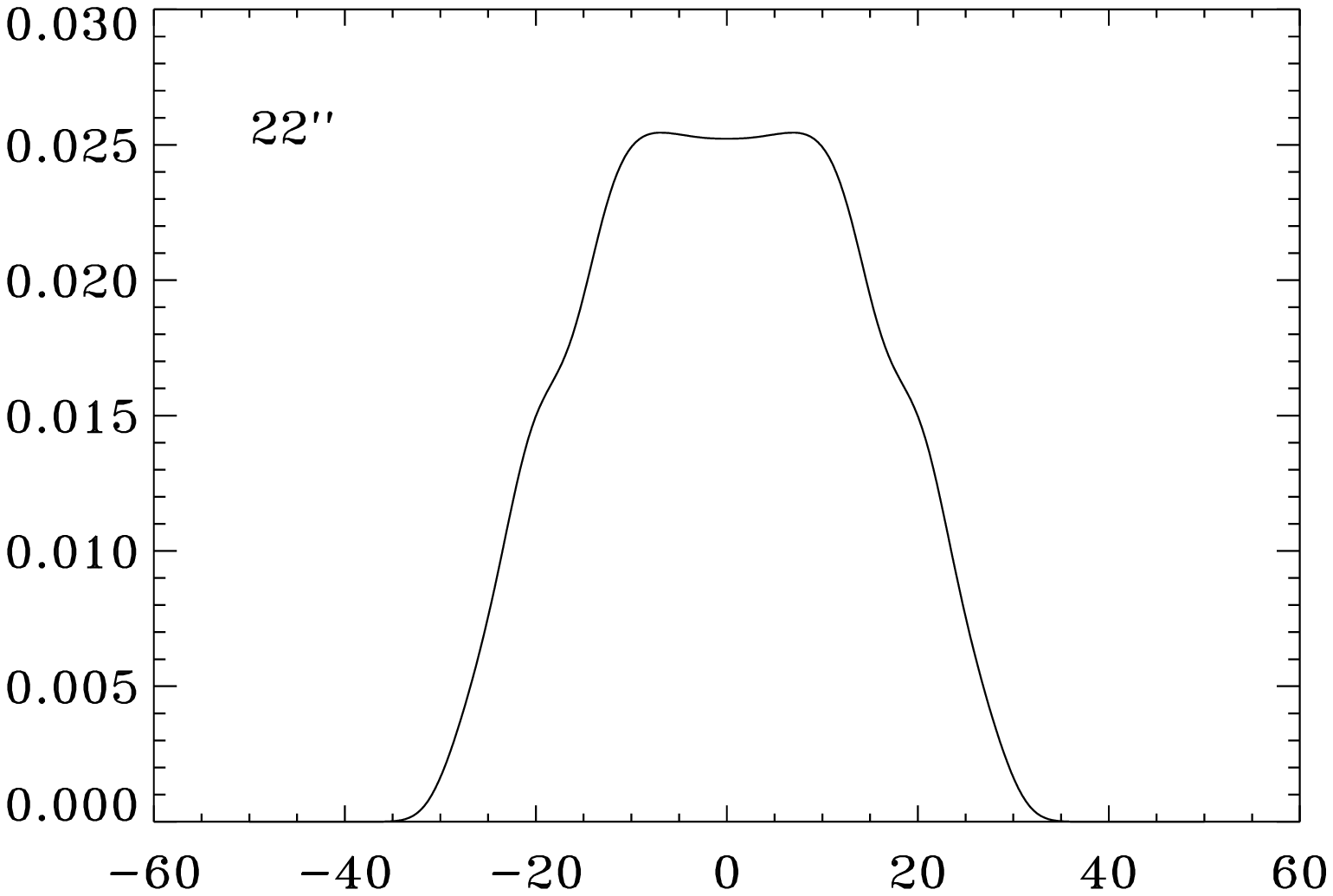}}
\caption{Line shapes for [\ion{O}{iii}]5007 at 11\arcsec, 15\arcsec, 18\arcsec,
and 22\arcsec\ {from the central star (for a distance of 1 kpc)}. The data has been convolved with a gaussian to achieve
10~km~s$^{-1}$ FWHM velocity resolution, and also 2\arcsec\ spatial resolution.}
\label{velpos-sim}
\end{figure*}

\subsubsection{Results of the simulation}

Figure~\ref{dvt-sim} shows plots of the density, velocity and temperature at
2550 years after the start of the simulation. The situation shown is the
periodic solution which develops after roughly 1200 years.  One sees the
result of a series of pulses, which are interacting and moving out. One
should realise that the faster parts of a pulse will run ahead of the slower
ones, and may be overtaken by the next pulse, leading to complicated
structures as a function of distance. The velocity  shows a `saw-tooth'
profile, which is still slightly accelerating outward.

Since our hydrodynamic model includes a photoionization model, we are
able to construct the I(5007)/I(4363) ratio from the local electron density,
temperature, and O$^{2+}$ density. We calculated the emission, turned it into
a three-dimensional distribution and projected this on the sky. In
Fig.~\ref{oiii-sim} we show the [\ion{O}{iii}]5007+4959
intensity, both as function of radius, and
along a radial cut through the image. One sees that the image shows the
typical emission characteristics of projected shells, just as {seen in} the
observations, cf.~BWH01. 

Figure~\ref{oiii-sim} also shows the electron temperature derived from the
I(5007+4959)/I(4363) ratio in the projected images, {as well as} on a cut through a
projected temperature map weighted with the H recombination emission. All
image data were smoothed with a gaussian of FWHM of 0\farcs 3. The
synthesized data can reproduce the electron temperatures found in the {\it
HST}\/ results, including the drop in temperature as function of radius. The
projection actually tends to hide the biggest variations in the line ratio,
which is due to the small filling factor of the areas of the highest
temperature. Comparing the cuts through the two temperature maps, one can see
how the [\ion{O}{iii}] ratio does give higher temperatures, as seen in the
case of the so-called `temperature fluctuations'. Note also how there actually is
a slight displacement between the high temperature peaks in both images. This
is because the H recombination-weighted image has a much stronger bias for
areas of higher density.

In Fig.~\ref{velpos-sim}, we illustrate the effect of the velocity variations
by showing synthesized line shapes at four different positions. The
velocities from the simulation were decreased  by 30~km~s$^{-1}$, and then
smoothed both along the velocity (FWHM 10 km~s$^{-1}$) and positional
(2\arcsec) directions in order to simulate the groundbased observations.
One sees how, at all four positions, the profiles are quite wide, consistent
with the observations.  {However, they are  not gaussian}; the fact that they orginate from
different shells is still visible. Since no line shapes have been published, 
it is hard to {say how} far our calculations  deviate  from the observations.

\section{Discussion}

The results of the simulation show that the interaction between mass loss
variations can reproduce shells which match the observed ones
morphologically, and to a large extent also kinematically and thermally. This
is the first attempt at {modelling properties of the shells, other than their 
appearance}. However, both the observational data and the modelling need 
improvement, and the results presented here should be viewed as a simple,
first attempt model. 

To start with the {observations}, narrowband images are not the best method for
measuring line ratios accurately, and a closer look at the kinematic data
would not hurt either. It would also be very interesting to see if similar
effects are found in other PNe with surrounding rings. Unfortunately, the
ideal observations would be difficult to carry out; to resolve the individual
shells, one needs the {\it HST}, but in order to do the spectroscopy on these
faint shells one needs a large aperture. Also, the velocity resolution of
the {\it STIS}\/ instrument on board {\it HST} is insufficient for a proper
analysis of the line shapes.

The shortcomings of the numerical model are obvious. For one, we were forced
to introduce an unrealistically high inflow velocity, and we set the
variations to be sinusoidal, which is hardly what one would expect. Ideally
one would like to follow the mass loss variations while the star is still on
the AGB, and let the different phases interact among themselves, and then
study the effect of photo-ionization when the central star evolves into a
post-AGB star. One would also like the mass loss from the star to vary with
its properties, so that a faster post-AGB wind starts to interact with the
inner parts of the AGB wind. This would solve the problem of the subsonic
inflow condition, as well as the question of the column density of the core
nebula. In short one would model the evolution of the circumstellar material
from the (late) TP-AGB until the PN phase. This has, for example, been done
by Steffen \& Sch\"onberner (2000), but their results do not address the
points raised in this paper. We plan to modify our simulations in order to
model mass loss variations during the AGB phase, and also their fate after
the star {evolves off the AGB}.

A further complication is the relation between the inner halo/ring area and
both the core nebula and outer halo. Compared to these two structures, the
inner halo seems an oasis of simplicity. The Chandra data have shown that the
core nebula consists of an inner wind-swept elliptical nebula, surrounded by
a (bipolar?) structure of lower ionization, pierced by FLIERs or jets along
the major axis. The chaotic structure of the outer halo also points to some
{dynamic interaction, altough this material was presumably lost long before
the inner halo and core nebula. The complete story of \object{NGC~6543} has not yet
been finalized.}

Still, it may be that the inner halo is indeed a relatively quiet area of
material lost just before the star turned off the AGB, in a period in which
the mass loss was variable in density and velocity, something which according
to some mass loss models is more likely to happen at the higher effective
temperatures and luminosities towards the end of the AGB (Simis, private
communication). The numerical study presented above shows that such a model
can explain the properties of the ``rings around the Cat's Eye''.

\section{Conclusions}

We analyze archival {\it HST}\/ [\ion{O}{iii}]4363 and 5007\AA\ narrowband
images and find an increased inner halo electron temperature of
$\sim$15,000~K for \object{NGC~6543}. This higher temperature is found in the same
region as the ``rings around the Cat's Eye'' reported by BWH01. This higher
electron temperature cannot be explained by photo-ionization models, and we
conclude that they must be produced by shocks. We show that, in principle, 
velocity and mass loss variations in the slow AGB wind, required for the
production and survival of the rings, would also be able to explain the
higher electron temperature in the inner halo, as well as the observed higher
velocity dispersion in that region.

\begin{acknowledgements}
Some  of the data presented in this paper was obtained from the
Multimission Archive at the Space Telescope Science Institute (MAST). STScI
is operated by the Association of Universities for Research in Astronomy,
Inc., under NASA contract NAS5-26555.

This research was supported in part by the KRF No. 2000-015-DP0445, by
the Korea MOST Grant Star No. Star~00-2-500-00, and by the KOSEF Grant
No. 2000-1-113-001-5 to KAO. SH is thankful to staff of the KISTI for
permission to use their Compaq GS320 supercomputer.  We also
thank the referee Dr. M. Peimbert, for a careful review and valuable
suggestions and Dr. A. Fletcher (KAO) for help with the preparation of
this paper.
 
The research of GM has been made possible by a fellowship of the Royal
Netherlands Academy of Arts and Sciences.  

\end{acknowledgements}

\end{document}